\documentclass[aps,pra,twocolumn,tightenlines,floatfix,showpacs,superscriptaddress,10pt]{revtex4-1}

\usepackage{upgreek}
\usepackage{epsfig}
\usepackage{tikz}
\usepackage{graphicx}
\usepackage{epstopdf}
\usepackage{fancybox}
\usepackage{makeidx}
\usepackage{mathrsfs}
\usepackage{appendix}
\usepackage{amsmath}
\usepackage{amsfonts}
\usepackage{natbib}
\usepackage[figuresright]{rotating}
\usepackage{floatpag}
\usepackage{natbib}
\usepackage{marginnote}
\usepackage{enumerate}

\newcommand{\xvec}{\mathbf{x}}

\newcommand{\kvec}{\mathbf{k}}

\newcommand{\svec}{\mathbf{s}}

\newcommand{\Rbb}{\mathbb{R}}

\newcommand{\I}{\mathcal{I}}
\newcommand{\nter}{\mathfrak{n}}
\newcommand{\mter}{\mathfrak{m}}

\begin{document}

\title{Memcomputing: Leveraging memory and physics to compute efficiently}

\author{Massimiliano Di Ventra}
\email{email: diventra@physics.ucsd.edu}
\affiliation{Department of Physics, University of California, San Diego, La Jolla, CA 92093}

\author{Fabio L. Traversa}
\email{email: ftraversa@memcpu.com}
\affiliation{MemComputing, Inc., San Diego, CA, 92130 CA}




\date{\today}

\begin{abstract}
It is well known that physical phenomena may be of great help in computing some difficult problems efficiently. 
A typical example is prime factorization that may   
be solved in polynomial time by exploiting quantum entanglement on a quantum computer. There are, however, other types of (non-quantum) physical properties that one may leverage to compute efficiently a wide range of hard problems. In this perspective we discuss how to employ one such property, {\it memory} (time non-locality), in a novel physics-based approach to computation: {\it Memcomputing}. In particular, we focus on {\it digital memcomputing machines} (DMMs) that 
are scalable. DMMs can be realized with non-linear dynamical systems with memory. The latter property allows the realization of a new type of Boolean logic, one that is {\it self-organizing}. Self-organizing logic gates are ``terminal-agnostic'', namely they do not distinguish between input and output terminals. When appropriately assembled to represent a given combinatorial/optimization problem, the corresponding self-organizing circuit converges to the equilibrium points that express the solutions of the problem at hand. In doing so, DMMs take advantage of the {\it long-range order} that develops during the transient dynamics. This {\it collective} dynamical behavior, reminiscent of a phase transition, or even the ``edge of chaos'', is mediated by families of classical 
trajectories ({\it instantons}) that connect critical points of increasing stability in the system's phase space. The topological character of the solution search 
renders DMMs robust against noise and structural disorder. Since DMMs are non-quantum systems described by ordinary 
differential equations, not only can they be built in hardware with available technology, they can also be simulated efficiently on modern classical computers. As an example, we will show the polynomial-time solution 
of the subset-sum problem for the worst cases, and point to other types of hard problems where simulations of DMMs' equations of motion on classical computers have already demonstrated substantial advantages over traditional approaches. 
We conclude this article by outlining further directions of study. 
\end{abstract}    

                          
\maketitle
\section{Introduction}\label{Intro}
{\it Computing with Physics $-$ } Computing is fundamentally a physical process. At the start and at the end of this process, it requires some agent (e.g., a person) to interpret the 
results of the computation. However, in between the end points of the computation, namely in between the input the agent supplies, and the output they interpret, {\it any} physical object or phenomenon can perform some type of computation. 

For instance, the planets in our solar system perform a well-defined motion with respect to the Sun. Without knowing, they 
{\it compute} their trajectory in a very complex environment, in fact, in the presence of the {\it whole} Universe surrounding them! Anyone equipped with a powerful enough telescope could determine the initial point of these trajectories (the input of the computation by the planets), and observe the final position of the planets after some time has elapsed (to read the output of the computation).
In the interval of time in between the observations, the planets have then ``calculated'' their orbits. Therefore, the telescope and the planets form some type of 
{\it computing machine}.

We could make similar considerations for {\it any physical system}. In a wide sense, {\it any} physical system performs some type of computation, whether it is easy or not for us to input the data at the beginning of such computation, or read them at the end of it. 

{\it Computing \`a la Turing $-$ }
Of course, this is not how, traditionally, we interpret computation. It is still commonplace to refer to the way in which Turing has formalized the  process of computing more than eighty years ago~\cite{36_turing}. Computing, \`a la Turing, is a mapping between a {\it finite} string of symbols into a {\it finite} string of symbols in {\it discrete} time~\cite{complexity_bible,computational_complexity_book}. The map itself, that transforms from the initial set of symbols to the final one, is called {\it transition function}~\cite{complexity_bible,computational_complexity_book}.

If we look back at the example of the trajectories computed by the planets we can definitely identify the transition function with the physical process (the laws of dynamics) that brings the planets from one point in space-time to another (without including the physical process that allows us to observe their initial 
and final positions). However, the computation the planets perform is {\it continuous} in time. Most importantly, 
since their initial and final positions, with respect to some reference frame, are real numbers, these numbers cannot be represented with a finite string of symbols: an 
arbitrary real number requires an {\it infinite} number of bits to be represented. Therefore, this type of computation, although physically possible, does not fall 
into the original Turing definition~\cite{36_turing}. A machine as the one represented by the telescope and the planets would be called {\it analog}, to distinguish it from the Turing one, which 
would be called {\it digital}.

Note that it is {\it not} the transition function that really distinguishes between a digital and an analog computer. Rather, it is the ability to read/write the 
output/input of the computation with finite means. If that were not the case, it 
would be physically impossible to build {\it any} digital computer. 

{\it Modern computers and scalability $-$ }
To make this important point clearer, consider our modern computers. In these, 
we manipulate finite strings of numbers (collections of 0s and 1s). However, we do this by representing these numbers as low (0) and high (1) current (or some other physical property) in actual 
{\it physical} devices, such as transistors; or patterns of magnetization in a
magnetizable material, or charges in capacitors, etc. \cite{Hennessy2011}.

Let us consider, for instance, transistors. Without going into details of how transistors work, it should not surprise anyone that it takes some {\it finite} time for any physical 
transistor to switch from a high current to a low current, and vice versa. Most importantly, since the dynamics of physical systems are {\it always} subject to noise and 
some other perturbations, there is no way that, at any given time, a transistor could be in a well-defined high current (representing the mathematical 1) or a 
low current (representing the mathematical 0). In other words, even if we neglected the time it takes a transistor to switch, it is a {\it physical impossibility} for it 
to represent the ideal, {\it mathematical} values of 0 and 1. Therefore, even our modern computers, which are classified as ``digital'', viewed as 
physical systems, are in reality ``analog''.

However, there is a very important difference between our modern computers and, say, the example of the ``planet analog computer''. Even though transistors cannot  represent a mathematical 1 or a 0 with absolute physical precision, the error one makes in assigning a given physical value of currents that represent those integers 
is {\it independent} of the size of the machine. Therefore, that error may be at most dependent on the ability to assign a 0 or a 1 out of a {\it single} transistor, 
but {\it it does not scale} with the number of transistors we put together.

Stated differently, although a physical realization of a (mathematically) ideal Turing machine does not exist, in practice 
our modern computers come very close to it: we can read and write outputs and inputs, respectively, of our modern computers with an error that is independent of their size. 
Without this important feature, which is not shared by true analog systems, our modern computers would suffer from {\it scalability} issues, in the sense that the bigger the size of the machine, the more {\it resources} (in time, space and/or energy) one 
would need to perform the computation with the same level of accuracy. 
With these caveats in mind, we may then rightfully call our modern computers ``digital''.

{\it The digital memcomputing paradigm $-$}
Why then, instead of trying to force a physical system (like our modern computers) to represent a (mathematical) Turing machine, don't we conceive of a completely different type of machines that take full advantage of appropriately designed physical systems, and yet allow us to read/write output/input digitally, 
namely with finite precision, or with an error that is independent of their size?

In this perspective article, we discuss such machines, that we have called {\it digital memcomputing machines} (DMMs)~\cite{dmm2}. DMMs are a subclass of a much larger 
class of {\it universal memcomputing machines} (UMMs)~\cite{UMM}, that include also analog ones~\cite{traversaNP}. We focus on their digital version, because, as we already mentioned, these are the ones that are easily scalable in terms of resources. 

UMMs have been shown to be Turing-complete, meaning that they can simulate any Turing machine~\cite{UMM}. Recently, it was also explicitly shown that they are  quantum-complete 
and reservoir-complete~\cite{ONUMM}, referring to the fact that they can simulate quantum computers as well as some type of recurrent spiking neural networks known as liquid-state machines~\cite{liquid_machine}.
Apart from these theoretical, albeit important conclusions, we have also shown that DMMs can be realized in practice by designing appropriate 
(non-quantum) non-linear dynamical systems. These dynamical systems have internal degrees of freedom (memory, hence the name {\it memcomputing}~\cite{13_memcomputing}). This allows us to engineer a new type of Boolean logic that is 
{\it terminal-agnostic}~\cite{dmm2}, namely one that does not distinguish between input and output. Therefore, while maintaining the digital structure of the input and the output (hence requiring finite means to read/write the output/input), DMMs built out of these new types of gates can {\it self-organize}, {\it collectively}, to solve very complex ({\it non-convex}) problems very efficiently. Loosely speaking 
then DMMs perform computation embedded in memory, employing all (or large chunks) of their fundamental units, at once. These are features that are generally attributed to the brain~\cite{kohonen_book}. 

{\it Memelements and more $-$ }The inspiration for the dynamical systems representing DMMs comes from electrical circuit elements with memory ({\it memelements})~\cite{09_memelements}. In general terms, a 
memelement is one that when subject to an input, $u(t)$, responds with an output, $y(t)$, with a generalized response function $g$ of the type
\begin{eqnarray}
y(t)&=&g\left(\tilde x,u,t \right)u(t) \label{Geq1}\\ \dot{\tilde x}&=&f\left(
\tilde x,u,t\right) \label{Geq2}
\end{eqnarray}
with $f$ some vector function of {\it internal state variables}, $\tilde x$, namely those variables that provide memory to the system, such as spin polarization, 
atomic position of defects, etc.~\cite{11_memory_materials}.

Memelements have been demonstrated to be good candidates not only for storing data \cite{08_strukov}, hence employed as alternative to current memory storage devices, but also to enable the possibility of building new generation of computational memories, i.e., memory devices that can perform basic computing tasks directly with and in memory \cite{borghetti_10,DCRAM,Pershin2015,Strukov-frontiers,lu-memristor, lu-sparse,lu-feature,indiveri}.   

However, it is important to stress that the machines we propose need {\it feedback} to operate as we desire~\cite{dmm2}. Therefore, {\it memelements alone are not enough}. One needs to add {\it active} elements as well, such as transistors, that are able to modify the state of the system ``on the fly'' according to specific rules to determine~\cite{dmm2}. Nonetheless, even 
these extra active components can be realized with standard electronics. Therefore, the machines we consider 
can be built in hardware with available materials and devices. In fact, they do not require more than standard complementary metal–oxide–semiconductor (CMOS) technology, if emulators of circuit 
elements with memory are employed~\cite{Pershin2010}. 

Importantly, since DMMs are non-quantum, their equations of motion can be simulated efficiently on our modern computers. Therefore, 
for some applications they can already deliver substantial advantages compared to standard algorithms for a wide range of non-convex (hard) problems of combinatorial/optimization type~\cite{exponential2017speedup,AcceleratingDL}. 

We will discuss the physics behind their power and provide as an additional example the polynomial-time solution of the search version of the subset-sum problem (which belongs to the NP-complete class~\cite{computational_complexity_book}). For this example we still employed {\it simulations} of DMMs on a single processor, showing 
once more that even without having been built in hardware yet, DMMs allow us to reap great benefits from just simulating them on our traditional 
computers. We finally conclude with future directions in the field.

\section{Leveraging Physics}\label{leverage}

\begin{figure*}[t!]
	\includegraphics[width=2\columnwidth]{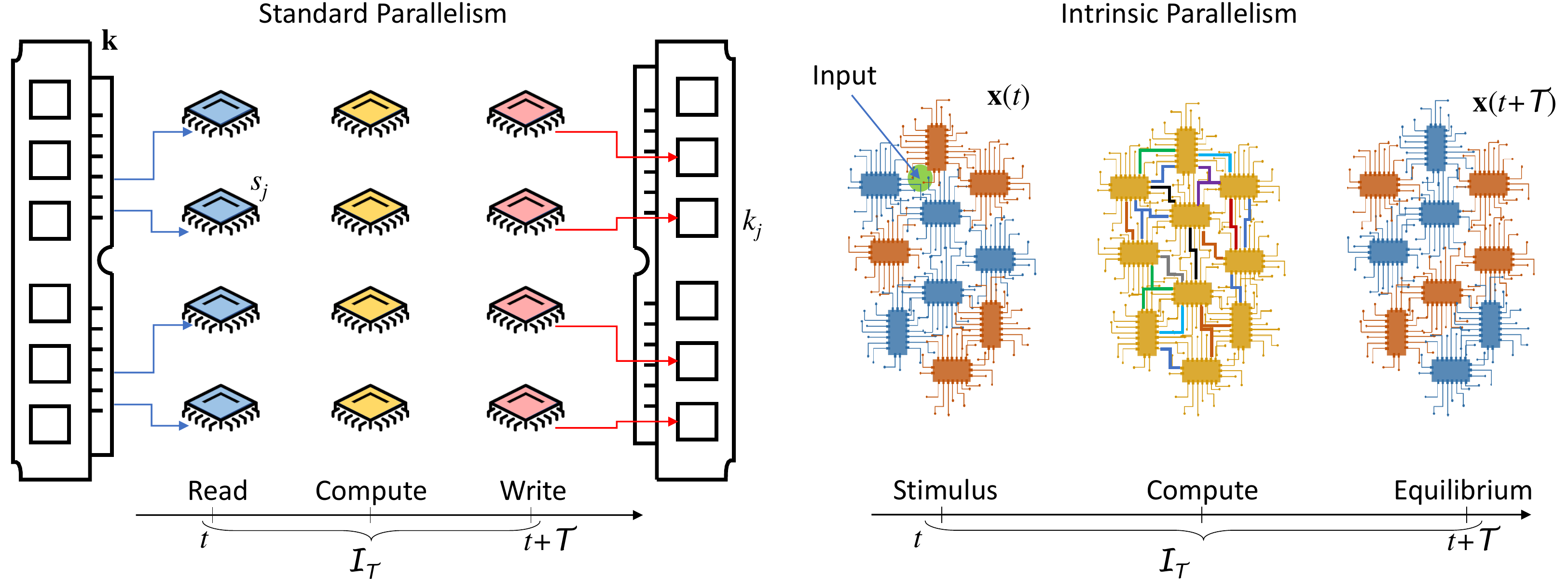}
	\caption{Left panel: representation of a standard parallel machine. Right panel: representation of an intrinsically parallel machine.}
	\label{figparallelvsintrisic}
\end{figure*}

{\it Dynamical systems for computation $-$ }We have anticipated in the Introduction that if we found a
physical system that, by solving a problem (say a combinatorial or optimization one), while preserving the digital (finite)
structure of the input and output, we may take advantage of its continuous
time (and space) power, and, at the same time, be able to interpret the results
of the computation with {\it finite} means. These computing machines would then be scalable. 

Here, we stress that ``time'' is not just ``counting steps'' of an algorithm, as algorithms are traditionally intended~\cite{complexity_bible,computational_complexity_book}.
Rather, it is a well-defined {\it physical} variable that labels the dynamics of the
system. Therefore, the correct framework in which we want to work is
that of {\it dynamical systems theory}~\cite{perko_01}.

In mathematical terms, a dynamical system is described by a (typically non-linear) differential equation of motion of the type (we consider only autonomous systems in our work)~\cite{perko_01}:
\begin{equation}
\dot \xvec(t) = F(\xvec(t)) \label{ODE},
\end{equation}
where $\xvec$ is an $n$-dimensional set of variables that defines the state of the system, and belongs to some $n$-dimensional space, $X \subset \mathbb{R}^n$, called phase space, and $F$ is a vector (called the flow vector field) representing the laws of temporal evolution of $\xvec$.

Equivalently, Eq.~(\ref{ODE}) can be written as 
\begin{equation}
T(t)\xvec(0)=\xvec(0)+\int_0^t F(\xvec(t^{\prime}))dt^{\prime}, \label{cr_semigroup}
\end{equation}
where $T(t)$ is a flow field. 

{\it Intrinsic parallelism $-$ } From either Eq.~(\ref{ODE}) or Eq.~(\ref{cr_semigroup}) it is easy to see why a physical dynamical system has already a feature, we call {\it intrinsic parallelism}, that is not shared by our standard ``parallel'' machines. To understand this important point, suppose a physical system described by Eq.~(\ref{ODE}) starts from a time $t$ in a (input) state $\xvec(t)$.
After an interval of time $\cal T$, it will be in a new (output) state  $\xvec(t+\cal T)$, as determined by Eq.~(\ref{ODE}). Therefore, Eq.~(\ref{ODE}) can be interpreted as a mapping, or (in computer science language) a 
{\it transition function}, $\delta$,~\cite{complexity_bible,computational_complexity_book} that relates the initial and final states, the input and the output, respectively, of the computation performed by this physical system. In this language, we can then write Eq.~(\ref{ODE}) in the form:
\begin{equation}
\delta(\xvec(t))=\xvec(t+\cal T)\label{deltaMM}.
\end{equation}

This equation may appear simple, but it represents a very profound and important characteristic for a machine whose computation is described by it. 
To see this, let us pause to consider the ``parallelism'' that is currently implemented in our modern ``parallel computers''. 

{\it Standard parallelism $-$ } Nowadays, to avoid increasing the clock frequency of computers (for heating issues), even laptops have multiple central processing units (CPUs) (or multi-cores), so parallel machines are becoming the norm rather than the exception. The definition of parallel machines we will use also includes some classes of parallel Turing machines, in particular the Cellular Automata and non-exponentially growing Parallel Random Access Machines \cite{Kozen_76,Fortune_78,Wolfram_84}.

Irrespective, let us consider a fixed number of (or, at most, a polynomially increasing number of) CPUs that perform some tasks in parallel. 
In this case, each CPU can work with its own memory cache or accesses a shared memory depending on the architecture. In any case, in {\it practical} parallel machines, all CPUs are {\it synchronized}. This means that each of them performs a task in an 
interval of time $\cal T$, which here means the synchronized clock time of the system of CPUs. At the end of the clock cycle, and {\it only} at the end of the clock cycle, all CPUs share their results, and follow up with the subsequent task. 

To make the comparison with Eq.~(\ref{ODE}) clearer, let us describe also these ``parallel machines'' within the mathematical framework of dynamical systems theory, an analysis that we have already outlined in~\cite{NANO_15}, and expand on here for clarity. 

Suppose that these machines have, say, $n_s$ CPUs with $n_s$ memory units. (The number of memory units may be different than the number of CPUs, but for simplicity of notation let us assume they are the same.) Consider the vector functions $\svec(t)=[s_1(t),...,s_{n_s}(t)]$ and $\kvec(t)=[k_1(t),...,k_{n_s}(t)]$ defining, respectively, the states of the $n_s$ CPUs, and the state of the slot $k_j$ of the total memory $\kvec$ allocated to be written by the CPU $s_j$. While the CPUs perform their computation, during each clock cycle $\cal T$ they operate {\it independently} of each other. Therefore, there are $n_s$ {\it independent} flow vector fields, $\phi^j_{\cal T}$ with $1\leq j \leq n_s$, describing the dynamics during the computation of the form 
\begin{equation}
(s_j(t+{\cal T}),k_{j}(t+{\cal T}))=\phi^j_{{\cal T}}(s_j(t),\kvec(t)),\;\;\;\;\;1\leq j \leq n_s, 
\end{equation} 
where $k_{j}$ is the memory unit written by the $j$-th CPU, and $\phi^j_{\cal T}$ is a function of its arguments, and which, in principle, could be different for the different CPUs.

Since the $j$-th CPU reads the memory $\kvec(t)$ at only the time $t$, and not during the interval $\I_{\cal T}=]t,t+{\cal T}]$, and it does not perform any change on it, apart from the unit $k_{j}$, the evolution (time-dependence) of the entire state during $\I_{\cal T}$ is completely determined by the set of 
$n_s$ {\it independent equations}
\begin{equation}
(s_j(t\rq{}\in \I_{\cal T}),k_{j}(t\rq{}\in \I_{\cal T}))=\phi^j_{t\rq-t}(s_j(t),\kvec(t))\label{PTMdyn},
\end{equation} 
with $1\leq j \leq n_s$. 

A quick comparison between Eq.~(\ref{PTMdyn}) with Eq.~(\ref{deltaMM}) shows that they describe {\it fundamentally different dynamics}. In each interval $\I_{\cal T}$ the $n_s$ CPUs {\it do not interact} in any way, and their dynamics are {\it independent}. We call then the dynamics described by Eq.~(\ref{PTMdyn}), {\it standard parallelism}. This is visually represented in Fig.~\ref{figparallelvsintrisic}, left panel.

On the other hand, the dynamics described by Eq.~(\ref{deltaMM}) are {\it collective}, namely {\it any} element of the vector $\xvec(t)$ is, at any given time, affected by the dynamics of {\it all} the other elements in the vector through their common flow vector field $F$. In other words, at any given time, 
{\it any} element of the machine is somehow ``aware'' of (or ``knows'') what the other elements are doing. 

As already mentioned, we call this {\it intrinsic parallelism}, to distinguish it from the standard parallelism implemented in our modern computers (see Fig.~\ref{figparallelvsintrisic}, right panel). This is, indeed, the power of {\it any} physical machine: it is the {\it physical interaction} among the different constituents of the machine that provides collective dynamics to the whole system.

Our goal is then to marry the {\it physical} power of a machine described by Eq.~(\ref{ODE}) with the {\it digital} structure of its input and output so that we need only {\it finite} means to read outputs and write inputs. To accomplish this we need to leverage more physical properties. 

\section{Leveraging memory}\label{LevMemory}

{\it Why memory? $-$ } Having established that a physical system described by Eq.~(\ref{ODE}) offers a type of parallelism that is not available in our modern computers, we have yet to answer the question: What type of physical interaction should we look for to compute efficiently? Or, to put it differently: How does ``knowing'' any other part of the 
system help computing a complex problem?

To answer this question, let us start by looking at an example that has attracted considerable attention in the past three decades: Quantum Computing~\cite{QI_bible}. The latter means computing by taking advantage of some features of Quantum Mechanics. In particular, in Quantum Mechanics, we have at our disposal a type of {\it spatial non-locality} known as {\it entanglement}~\cite{QI_bible}. Entanglement allows a quantum machine with such characteristic to have its elements ``correlate'' with each other at very long distances, as if the whole system were ``rigid'': a perturbation in (measurement of) one of its parts, 
would be immediately felt in other parts arbitrarily far away. Entanglement then realizes an {\it ideal long-range order}, one in which correlations do not decay 
spatially.  

It is this type of interaction that allows a Quantum Turing Machine (QTM) to solve a non-deterministic polynomial (NP) problem, such as 
factorization, efficiently, albeit probabilistically~\cite{Shor_1}. Unfortunately, a QTM has not been shown to efficiently solve other, more difficult  
problems, such as NP-complete problems, although quantum computers (that support entanglement) are now being engineered to find the ground state of some quantum Hamiltonians~\cite{q_hamiltonian}. However, in order to take full advantage of entanglement, these machines need to work at extremely low (cryogenic) temperatures, with substantial increase in hardware complexity, and difficulty in scaling them up to large size~\cite{Ladd2010}. In addition, since the Hilbert space of a quantum system 
typically scales exponentially with the size of the system (e.g., the number of its elementary units), quantum machines {\it cannot} be simulated on our modern 
computers efficiently.

It would then be desirable to have some type of long-range order (that allows distant parts of a machine to correlate with each other efficiently), without recurring to the entanglement of Quantum Mechanics.  In fact, long-range order is quite a common physical feature. It is shared by many systems and emerges in a
wide-variety of phenomena and structures such as continuous phase transitions~\cite{Goldenfeld}, self-organized criticality \cite{Bak1988}, and complex networks \cite{BOCCALETTI2006}, to name a few. Therefore, it should not come as a surprise if it
emerges in appropriately-designed dynamical systems. 

Typically, classical systems have {\it spatial} interactions that are {\it local/short ranged} (if we exclude gravitation and note that Coulomb interactions 
in condensed matter systems are generally screened). Therefore, spatial non-locality in non-quantum systems seems difficult to achieve from only local interactions, 
unless we work, say, at a phase transition, or somehow engineer the system to have correlations that are {\it scale-free} (decay as a power law). 

Instead, we point out an obvious fact: {\it any} physical system (whether classical or quantum) supports some level of {\it memory} (time non-locality)~\cite{kubo1957statistical}. 
In fact, it can be shown that {\it any} system (whether passive or active) with memory (any memelement) can be compactly written as in Eqs.~(\ref{Geq1}) and~(\ref{Geq2}), starting from its 
microscopic dynamics~\cite{13_properties}.  
This is because no physical system can respond instantaneously to a perturbation. (Of course, in many cases this memory is difficult to detect, but this is beyond the point we are making here.)

{\it Memory promotes spatial non-locality  $-$ } Memory then seems a good ingredient to exploit, if it does lead to long-range correlations. This has already been shown in, e.g., complex networks, where scale-free properties 
(such as a power-law distribution of the degree connectivity in the network) can emerge by allowing only time non-locality~\cite{caravelli,Caravelli2017}. That this is the case, can be intuitively understood from an example taken from the natural world: how ants find the shortest path to a food source from their nest \cite{Blum2005}.

In order to find food, a few ants are initially randomly dispatched out from their nest. These ants randomly scout the nest surroundings, and, while doing
so, release a chemical substance (pheromone) that can be detected by other ants, and which decays in time. In simple words, these ants leave a ``memory
trace'', with this memory decaying in time. Now, if we assume that the decay time of the pheromone memory is, on average, the same for all ants, after some time has elapsed, the longest the path traced by an explorer ant, the less pheromone it would contain, compared to another, shorter path traced by another ant. Other ants then exiting the nest would be attracted by the strongest pheromone scent, and hence would be drawn towards the shortest path, rather than the other paths. 

{\it Collective behavior and self-organization $-$ } 
While going through the shortest path, these ants then reinforce it by further releasing pheromones, so that, ultimately, the shortest path (or one close  
to it) is {\it collectively} chosen
by the colony. Note that the ant colony uses two main ingredients to solve this path (network)
optimization problem: memory and ``collective behavior''. A single ant does not solve the problem. It is the cooperative/collective
behavior of many ants that accomplishes the task.

If the pheromone memory did not exist, ants being non-quantum objects
(and assuming they do not interact with other ants in any other physical
way), have no means to ``correlate'' with each other at long distances, and
hence solve the shortest-path problem. In fact, due to the memory trace they leave, ants {\it self-organize} into the solution, namely in a total {\it unsupervised} way (without an external agent guiding them), they find the solution to the shortest path. This same self-organizing behavior has been shown to occur also in networks of memristors (resistors with memory)~\cite{13_self_organization,antcolony}. 

The above example then clearly shows that spatial correlations (space non-locality) among physical (and non-quantum) systems can emerge from time non-locality (memory) alone, even if the physical constituents themselves interact locally. In addition, collective dynamics with memory leads naturally to the phenomenon of {\it self-organization}. 

Of course, 
for computing purposes, we are not interested in any type of spatial correlations. In fact, most of the time, correlations between physical systems decay 
exponentially~\cite{Goldenfeld}. We would like instead those correlations to decay at most as a power law ({\it scale-free} behavior)~\cite{Goldenfeld}. Ideally, we would like 
those correlations not to decay at all, as in the case of entanglement of Quantum Mechanics ({\it ideal scale-free} behavior)~\cite{QI_bible}. This way, distant parts of our 
(non-quantum) machine may correlate and self-organize easily into the solutions of the problems they are designed to tackle. In order to realize a machine 
that has the above features, we first need to replace the standard Boolean logic framework with a new one. 

\section{A new logic framework}\label{newSOLGs}
Memory and self-organization allow us to address the last aspect of the computation we are after: read and write outputs and inputs, respectively, with {\it finite} precision. 
Since finite precision simply translates into expressing a problem in {\it binary} (Boolean) format, this means that we focus here on all those problems that are naturally written in Boolean format, such as combinatorial/optimization problems~\cite{complexity_bible,computational_complexity_book}. We will discuss in the conclusions how to apply memcomputing to other 
types of ``continuous-variable'' problems. The question now is how to utilize a physical system described by Eq.~(\ref{ODE}) to solve a Boolean problem. 

{\it Example: prime factorization $-$ } To this end let us consider an example: factoring a number into primes, a problem that is believed to belong to the NP class (albeit it is not NP-complete)~\cite{computational_complexity_book}. Suppose you are given an integer $n$ that, for simplicity, can be factored into only two prime numbers, say $p$ and $q$, $n=pq$. 
Due to the fundamental theorem of arithmetics \cite{Apostol2010}, if the number $n$ can be factored in two prime numbers, those are unique. 

All these numbers can be expressed in binary representation as
$n=\sum_{j=0}^{N}n_{j}2^{j}$, $p=\sum_{j=0}^{N-1}p_{j}2^{j}$ and $q=\sum_{j=0}^{\lceil N/2\rceil -1}q_{j}2^{j}$, where, $n_j$, $p_j$, and $q_j$ are the 0s and 1s representing the respective integers; $N=\lfloor\log_{2}n\rfloor$, $\lfloor X\rfloor$ is the floor function that rounds the elements of $X$ to the nearest integer towards $0$, and $\lceil X\rceil$ is the ceiling function that rounds the elements
of $X$ to the nearest integer towards $\infty$.

Since we have assumed that $p$ and $q$ are primes, then we can choose $p_{0},\,q_{0}\neq0$, which guarantees that $p$ and $q$ are not divisible
by $2$. This means that $n_{0}=1$ and we also set $n_{N}=1$. This implies that $n$ has the shortest binary representation.

Let us then assume we know $p$ and $q$. If we multiply them to get $n$, we would obtain at each step of the computation the following remainders, $r_{j}$, of the above arithmetic operation:
\begin{align}
r_{0} &  =p_{0}q_{0}-n_{0}=1\times1-1=0;\\
r_{j} &  =\sum_{k=0}^{j}p_{j-k}q_{k}+\frac{r_{j-1}}{2}-n_{j},\;\;j=1,...,\lceil N/2\rceil-1;\\
r_{j} &  =\sum_{k=0}^{\lceil N/2\rceil-1}p_{j-k}q_{k}+\frac{r_{j-1}}{2}%
-n_{j},\;\;j=\lceil N/2\rceil,...,N-1;\\
r_{N} &  =\sum_{k=0}^{\lceil N/2\rceil-1}p_{N-1-k}q_{k+1}+\frac{r_{N-1}}%
{2}-1,
\end{align}
that must satisfy
\begin{align}
r_{j} &  \geq0\,, &j=1,...,N-1\,,\nonumber\\
r_{j} &  =0\,,\,\,\operatorname{mod}2&j=1,...,N-1\,,\nonumber \\
r_{N} &  =0\,.
\end{align}

This is standard arithmetics. However, it lends itself immediately to a Boolean representation. In fact, by looking at the truth tables of the XOR and AND gates \cite{Givant2008}, we easily recognize the following mapping between the arithmetic operations of sum, $\sum$, and product $\times$, with the 
corresponding logic gates: $\sum \rightarrow \textrm{XOR},\textrm{AND}$, and $\times \rightarrow \textrm{AND}$. 
This means that the entire arithmetic operation of factoring an integer can be implemented as a Boolean problem. In fact, {\it any} combinatorial or optimization 
problem can be written as a {\it Boolean circuit}~\cite{computational_complexity_book}. 

\begin{figure}
	\centerline{
		\includegraphics[width=\columnwidth]{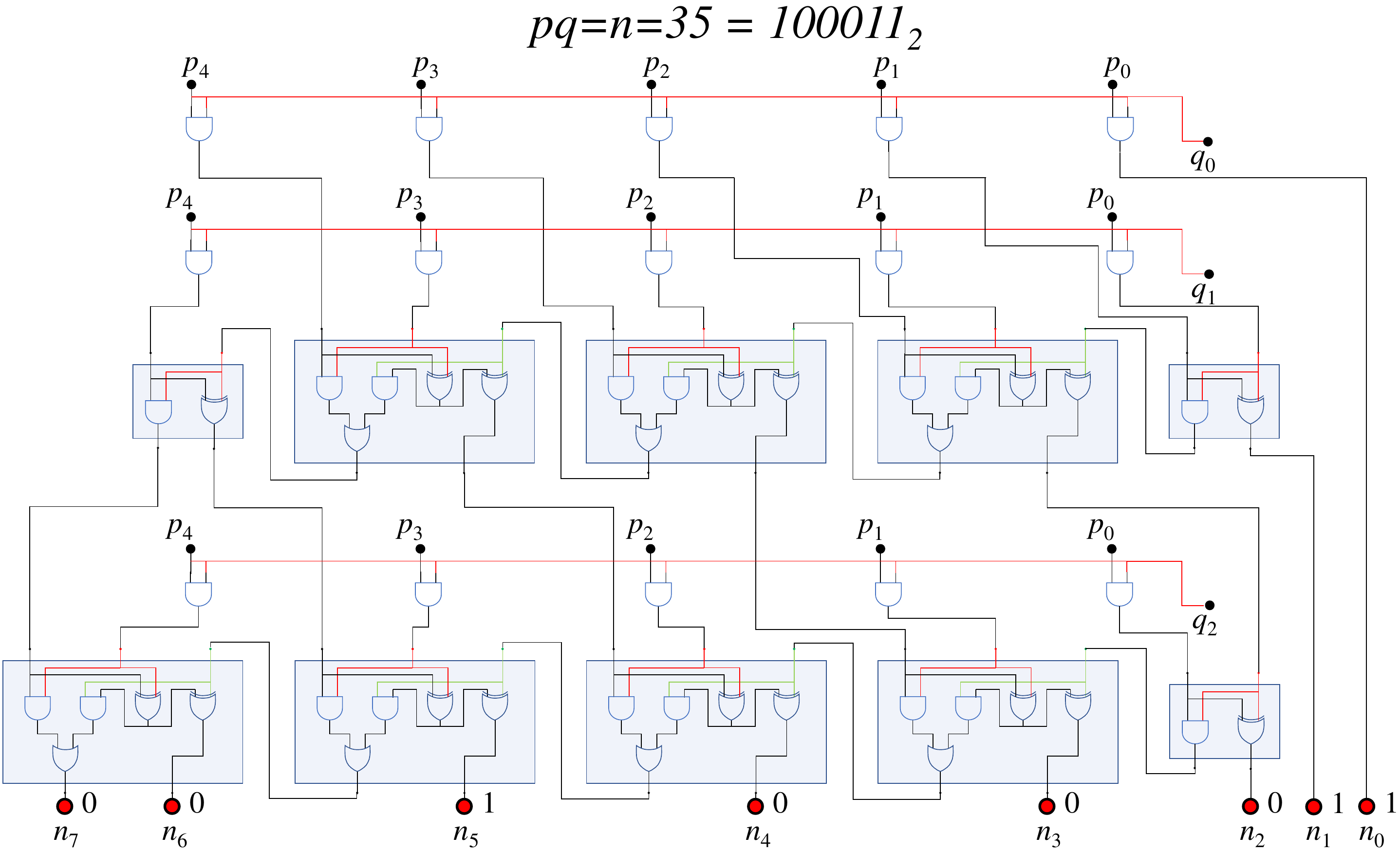}}
	\caption{\label{Boolfactor} A possible Boolean circuit that multiplies two integers $p$ and $q$ to give $n=35=(100011)_2$ (in the little-endian notation).}
\end{figure}

Now, the Boolean circuit that summarizes the above arithmetic operation is {\it not} unique. One could use different Boolean gates and circuits to accomplish the same task. This is because we could use different logic gates as a basis of Boolean logic. For instance, the pairs $\{$AND, NOT$\}$ or $\{$OR, NOT$\}$ are two sets of Boolean gates that can be used to represent {\it any} Boolean gate \cite{Givant2008}. The gates NOR (not-OR) or NAND (not-AND) represent another complete (singleton) basis set. As an example of a Boolean circuit factoring a number, we show in Fig.~\ref{Boolfactor} one of the many possible circuits 
that accomplishes the multiplication of $p$ and $q$ to obtain the number $n=35=(100011)_2$ (in the little-endian notation). 

{\it Self-organizing logic $-$ } A machine that, given $p$ and $q$, uses that circuit, would spit out $n$. Of course, to solve prime factorization we need to work {\it in reverse}: given $n$, we want our machine to find $p$ and $q$. 

\begin{figure}
	\centerline{
		\includegraphics[width=.8\columnwidth]{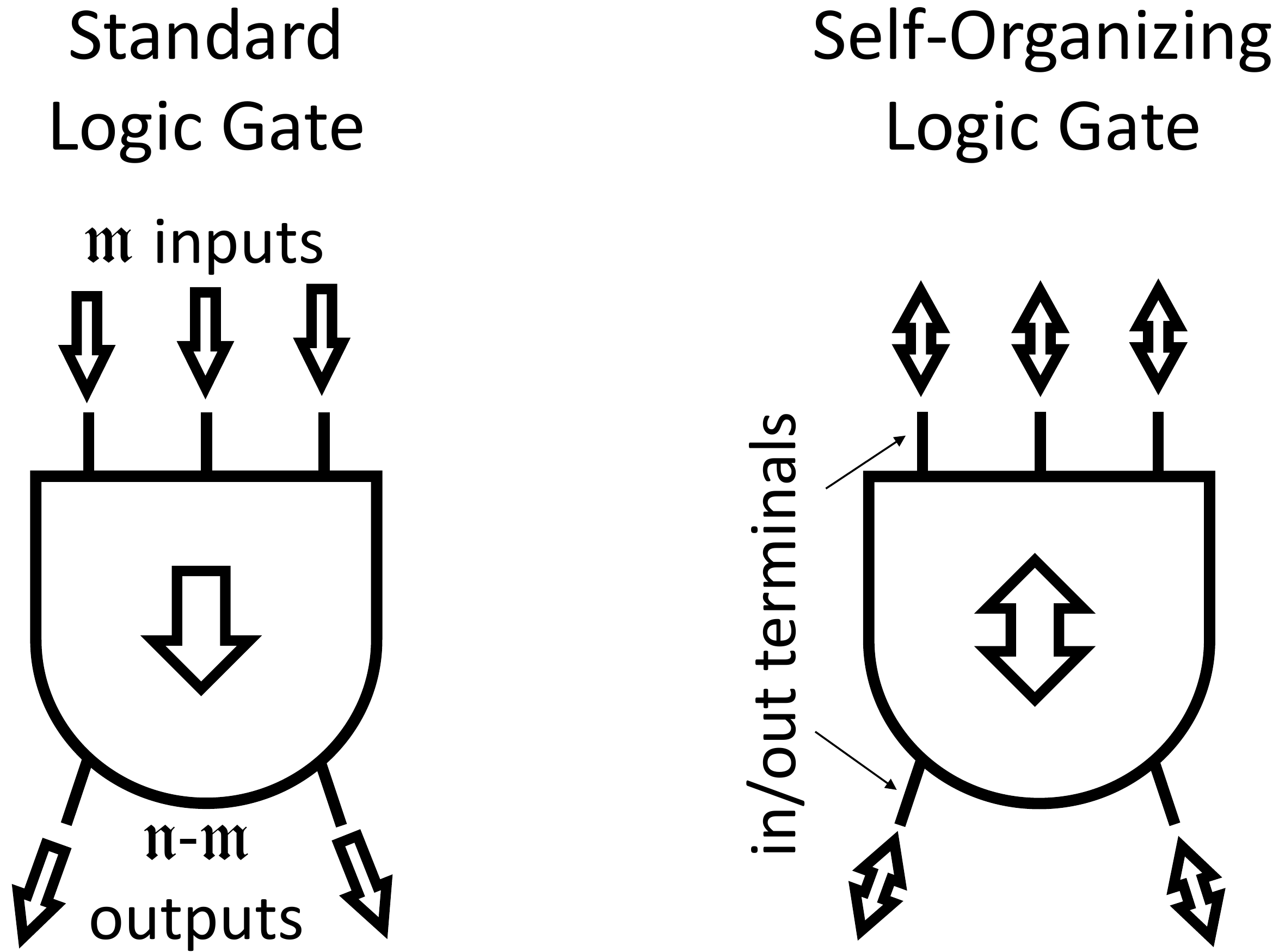}}
	\caption{\label{DMM_SOLG-sketch}Left panel: sketch and symbol of a standard $\nter$-terminal logic gate with $\mter$ inputs and $\nter-\mter$ outputs. Right panel: sketch and symbol of the corresponding self-organizing logic gate.}
\end{figure}

It is obvious that if we used standard logic gates, they would not allow us to solve the problem. A standard Boolean gate is a mapping that receives truth values of some {\it input} variables (we may also call them {\it terminals}), and spits out the truth values of some {\it output} variables/terminals (see Fig.~\ref{DMM_SOLG-sketch}, left panel). In other words, standard Boolean logic is {\it sequential}: given truth values of some {\it input} variables (terminals), it provides a truth value of {\it output} variables (terminals). 

However, we have discussed in Sec.~\ref{LevMemory} that memory (time non-locality) may promote spatial non-locality, so that physically distinct parts of the same system can effectively communicate with each other, even if the interactions are local. Let us then consider a Boolean gate as a {\it physical} system, and its ``input'' and ``output'' terminals as physically distinct {\it states} of that system, which we may intuitively think of as being located at distinct regions of 
space. 

With this new point of view in mind, we can {\it design} gates in such a way that, {\it regardless} of the terminal to which the truth value is assigned --in what we have called so far ``output'' terminal or ``input'' terminal-- the logic proposition is {\it always satisfied}. This is a type of {\it terminal-agnostic logic}. Note that this is {\it not} the same as inverting the logic in a one-to-one sense, as, e.g., in Toffoli gates~\cite{toffoli}. The latter ones have the same number of input and output terminals, hence they can define a bijection. Our gates generally do not. 

Take for instance, the AND gate. There are two ``input'' terminals, and only one ``output'' terminal. Hence, this gate cannot be inverted in a one-to-one sense.  
However, we can always {\it demand} that, if the ``output'' terminal 
is fixed at the truth value 1, the two ``input'' terminals need to be both 1. Instead, if the ``output'' terminal is 0, the two ``input'' terminals need to have one of these three possibilities, (0,0), (1,0), (0,1), in order for this gate to be {\it logically consistent}. That is all we ask of our gates. In this way, the distinction between input and output is no longer necessary, i.e., signals can go in and out at the same time at {\it any} terminal, resulting in a (albeit non-linear) ``superposition'' of input and output signals as depicted in the right panel of Fig.~\ref{DMM_SOLG-sketch}. 

One immediately recognizes that in order to have this extra feature, not present in standard Boolean gates, we need to add extra {\it degrees of freedom} to the system. This way, the system must be able to {\it adapt} or {\it self-organize} to any value of the terminal, 
and to do so, it needs extra ``room to maneuver''. 

As already mentioned, we then construct these gates by adding time non-locality, namely a dependence on internal state variables. In practice, this means using memelements (Eqs.~(\ref{Geq1}) and~(\ref{Geq2})), or their emulators, and other standard circuit elements (active and passive), appropriately designed to satisfy the correct logical propositions of the gate. We encode the terminals' truth values with voltages (or currents) of these circuits, e.g., $+1$V representing the logical 1, 
while $-1$V the logical 0~\cite{dmm2}. 

\begin{figure}
	\centerline{
		\includegraphics[width=.8\columnwidth]{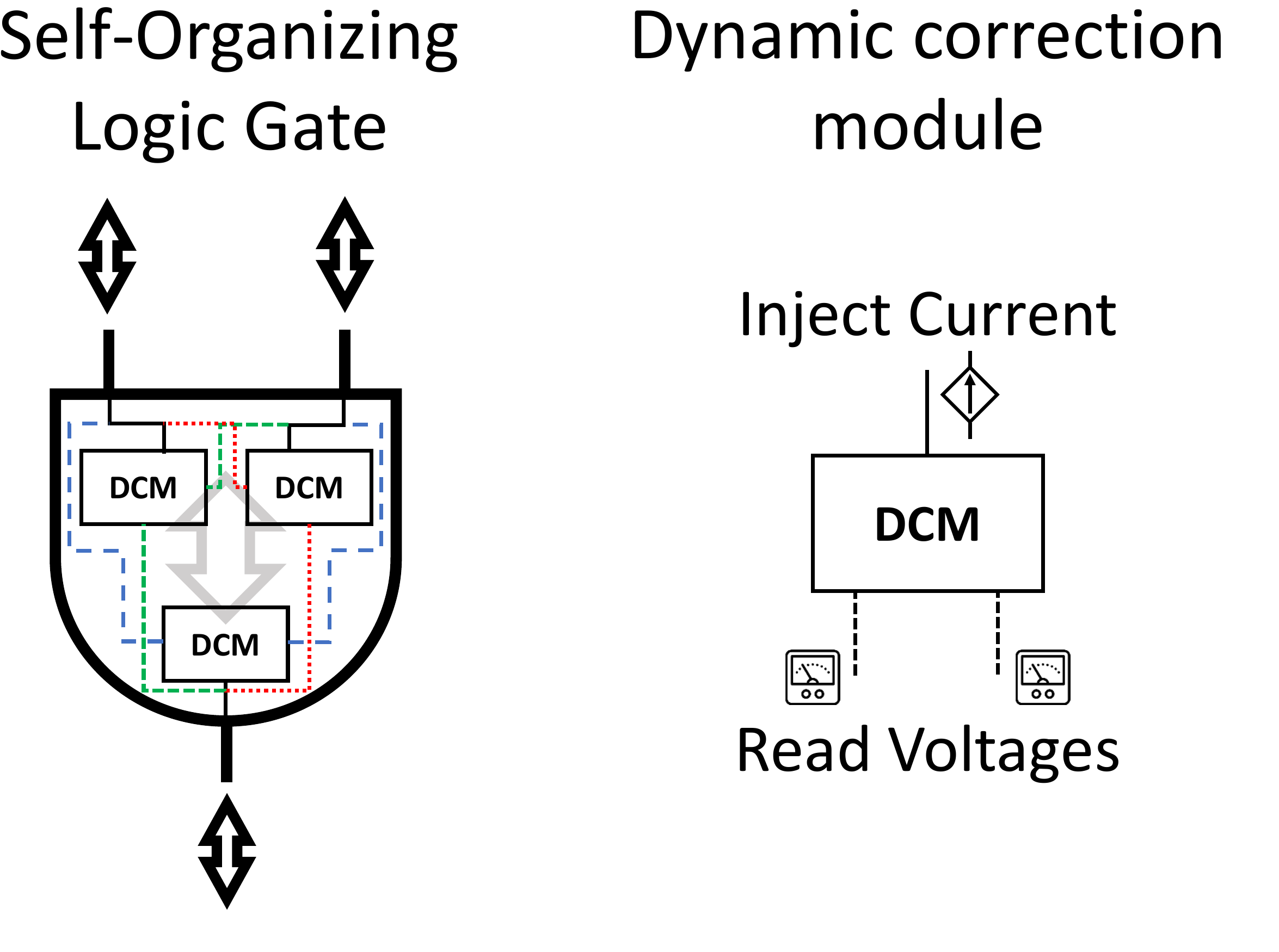}}
	\caption{\label{DMM_DCM-sketch}Left panel: sketch and symbol of internal modules of a 3-terminal self-organizing logic gate. Right panel: sketch and symbol of the dynamic correction module.}
\end{figure}

The  active elements of the circuit have the role of {\it dynamically correcting} the gate state, if the latter is in 
a wrong configuration. We call these {\it dynamic correction modules} (DCMs)~\cite{dmm2}, see Fig.~\ref{DMM_DCM-sketch} for a sketch of DCMs. For instance, in the gates we have proposed in Ref.~\onlinecite{dmm2} the DCM dynamically reads the voltages at the terminals of the gate, and injects a large current when the gate is in an inconsistent configuration, a small current otherwise. 
We re-iterate that this task cannot be done by only passive circuit elements, whether they have memory or not. It requires active elements. 

Now, since these gates are described by dynamical systems, they will 
dynamically reach their consistent logical proposition according to the {\it initial conditions} they are in, namely not just the initial Boolean truth value the terminals 
have, but also what value the internal state variables start from in the {\it attraction basins} of each equilibrium point. In other words, we need to assign the value of $\xvec (t)$ of Eq.~(\ref{ODE}) at $t=0$. We call the objects having these logical and dynamical properties \textit{self-organizing logic gates} (SOLGs), see Fig.~\ref{DMM_SOLG-sketch}. We name the circuits built out of SOLGs, {\it self-organizing logic circuits} (SOLCs). These are an actual physical 
realization of DMMs~\cite{dmm2}. 

Of course, DMMs could be realized in other ways. For instance, one may design SOLGs using optical devices. The connections between these SOLGs to generate SOLCs may also be realized using optical means. We focus here only on their electronic realization, since it is the easiest to build in hardware (and arguably the only way for very-large-scale integration (VLSI) realizations) and simulate in software.

\section{The mathematical requirements of SOLCs}\label{MathSOLCs}

So far, we have shown how to transform a Boolean problem into a physical problem, so that we can solve it also {\it in reverse}. This allows 
us to invert so-called ``one-way functions''~\cite{computational_complexity_book}, namely those problems, such as factorization, that are easy to solve 
in one direction (from $p$ and $q$ one easily gets $n=pq$), but not the reverse (from $n$, find $p$ and $q$). By doing so, we have maintained the {\it digital} structure of input and output (the terminals of the SOLGs can be written/read with finite means). We are then on our way to exploit the power of a physical system, while maintaining the {\it scalability} of the computing machine. However, we have quite a bit of freedom to engineer SOLGs (and corresponding SOLCs) in such a way that when they are built in hardware or simulated 
in software they perform the tasks we require of them, such as solving the Boolean problems they have been designed to tackle. 

First of all, it is clear from Eqs.~(\ref{Geq1}) and~(\ref{Geq2}) (and also the equations of the dynamic correction modules~\cite{dmm2}), that Eq.~(\ref{ODE}) turns out to be a {\it non-linear} equation of all the state variables (voltages, currents and memory variables of the whole circuit) lumped into the state $\xvec(t)$. However, not all non-linear dynamical systems with memory are good for the job. This is because a general non-linear dynamical system of the type~(\ref{ODE}) may have several 
attractors with unwanted features, hence it may not solve the problem we are after. To make this point clear, let us consider the following problem, known as 
subset-sum problem (belonging to the NP-complete class)~\cite{complexity_bible,computational_complexity_book}. 

{\it Example: the subset-sum problem $-$ } Consider a set $G$ of $N$ integers ($N\in \mathbb{Z}$), each represented with $p$ bits (precision of each number). Consider another integer $s\in \mathbb{Z}$. Question: is there a subset of $G$ whose sum is $s$? This problem may have no solution, one solution or many. In fact, if there is a solution, 
we would like to know the integers that do satisfy the sum (``search version'' of the problem)~\cite{complexity_bible,computational_complexity_book}. 

As for the factorization problem, we can build a Boolean circuit that represents this problem. We then replace the gates of that circuit with our SOLGs. The whole circuit can be represented 
by an equation of motion of the type~(\ref{ODE}) with the state variables, $\xvec(t)$, again describing voltages at the SOLGs' terminals, currents in the circuit, and internal state variables representing memory~\cite{dmm2}. We then let the dynamical 
system ``run in reverse'' to find the numbers (if they exist) that sum to $s$. 

{\it The choice of dynamical systems $-$ } We face several problems if we pick some arbitrary 
dynamical system with memory to accomplish this task. First of all, how do we encode the solution(s) to this problem? The easiest (and most natural) way 
to do so is to encode the solution(s) of the problem into the steady-states (equilibrium points) of Eq.~(\ref{ODE}), namely we read the solution (the voltages) at the appropriate 
SOLGs encoding the output of the calculation, when all SOLGs 
have satisfied their logical proposition (equilibrium points of Eq.~(\ref{ODE})). 

If we follow this path, however, we need to avoid the 
system falling into a periodic orbit from which it will never exit, or end up into a strange attractor (chaotic behavior)~\cite{Synergetics}. On top of these requirements, 
we need the system to reach the equilibrium points {\it exponentially} fast. Finally, if we increase the size of the problem (say, the number of elements $N$ in the 
set), hence the number of SOLGs in the circuit, the convergence rate to equilibrium points better scale at most {\it polynomially} with size, or we have not 
gained much with respect to traditional (algorithmic) approaches. (In fact, we want {\it all} the resources to scale polynomially, not just time.)

{\it Point dissipative systems $-$ } In Ref.~\cite{dmm2} we 
have suggested a set of dynamical systems with memory that accomplishes all these properties. The main, most important, ingredient is that the dynamical 
systems we suggested are {\it point dissipative}~\cite{Massatt1981,Massatt,hale_2010_asymptotic}. These are special dynamical systems that support a compact {\it global attractor}. This implies that 
all trajectories of the system are {\it bounded} and will ultimately end up into the global attractor, {\it irrespective} of the initial conditions. 

From a mathematical point of view, this means that the flow field $T(t)$ 
in Eq.~(\ref{cr_semigroup}) can be written as the sum~\cite{dmm2,hale_2010_asymptotic}
\begin{equation}
T(t)=U(t) + S(t) \label{split}.
\end{equation}
The functions $U(t)$ and $S(t)$ are two vector fields that, however, must be chosen with completely different dynamical behavior. In fact, a point dissipative system requires that for $S(t)$ there exists a continuous function $k:\Rbb^+\rightarrow\Rbb^+$ such that $k(t,r)\rightarrow0$ as $t\rightarrow\infty$ and $|S(t)\xvec|<k(t,r)$ if $|x|<r$, 
with $r$ a positive constant~\cite{hale_2010_asymptotic}. This means that in the long-time limit only the function $U(t)$ has any effect on the dynamics of the state. 

Now, to fully exploit this property 
we design $U(t)$ appropriately, namely we {\it choose} $U(t)$ to describe a {\it globally passive circuit}~\cite{Passive}. This means that the equilibrium points of $U(t)$,  
hence of the total flow field $T(t)$, are reached exponentially fast. If we make this choice, we have an additional benefit: in the presence of equilibrium points, $U(t)$ (and consequently $T(t)$ or $F(t)$) {\it cannot} support either periodic orbits or chaos~\cite{Passive,noperiod}. 

As an example, the functions $U(t)$ and $S(t)$ can be read from Eqs.~(52)-(55) and Eqs.~(56)-(59) of Ref.~\onlinecite{dmm2}, respectively. In fact, in Ref.~\onlinecite{dmm2}, we have shown that, for that particular choice of dynamical systems, there is a constant $\xi>0$ such that $|S(t)\xvec|<e^{-\xi t}$.

We then conclude that one can indeed engineer dynamical systems representing SOLCs with appropriate mathematical properties that {\it guarantee} to find the solution(s) of the given Boolean problem, if they exist. If there are multiple solutions, the system will find one of them according to the initial conditions assigned. Typically, one solution is required. If all of them are needed, one can simply add extra constraints to the circuit that take into account 
the solution(s) already found, and repeat this process till all solutions are found. Finally, 
if no solution exists, by knowing the scalability of the corresponding problem in terms of its size, one can check if the system has reached equilibrium 
or not: if it does not within the expected time, then no solution exists~\cite{dmm2}. 

{\it Properties of SOLCs $-$} Let us then summarize the mathematical properties of the SOLCs we have introduced in Ref.~\onlinecite{dmm2}. These are the properties that any other type 
of dynamical system has to satisfy in order to solve a given Boolean problem efficiently (namely with polynomial resources):\\
 
\begin{enumerate}[i)]
	\item Design SOLCs whose equations of motion are point dissipative.
	\item Design them so that there cannot be equilibrium points other than solutions of the given problem at hand.
	\item For each size of the problem, equilibrium points are reached exponentially fast from all points 
	in their attraction basin.
	\item The convergence rate scales at most polynomially with input size.
	\item SOLCs' resources grow at most polynomially for problems whose solution tree grows exponentially.
	\item In the presence of equilibria (solutions) there cannot be periodic orbits or chaos.
\end{enumerate}

Note that the SOLCs we have proposed in Ref.~\cite{dmm2}, do satisfy all criteria i) - vi), including the last one (see Refs.~\cite{noperiod} and~\cite{no-chaos}). Once SOLCs with these properties have been designed, we can follow this procedure to tackle any Boolean problem: \\

\begin{enumerate}[i)]
\item Construct the Boolean circuit that represents the problem at hand (this circuit is not unique). 
\item Replace the traditional (uni-directional) Boolean gates of this circuit with SOLGs. The electronic circuit built out of these SOLGs can be described by {\it non-linear} ordinary differential equations with memory of the type~(\ref{ODE}).
\item Feed the appropriate terminals with the required ``input'' of the problem (e.g., the number that needs to be factored). 
\item Build the corresponding SOLC in hardware or simulate its differential equations in software. 
\item Find the equilibrium (steady-state) points of the dynamical system, which encode the solution to the problem. 
\end{enumerate}

{\it Combinatorial vs. optimization problems $-$ } A clarification is in order. In the case of {\it combinatorial} problems (such as factorization and subset-sum) the SOLCs designed to tackle them, {\it do} solve them, 
if solutions are present. In other words, we {\it can} guarantee that if the combinatorial problem has a solution, the SOLC, designed for that problem, will find it 
(see also discussion in Sec.~\ref{Topology}). It is also easy to check if the solution is the correct one: we can do it in polynomial time once we have a candidate solution. 

On the other hand, when we deal with {\it optimization} problems (especially those in the NP-hard class), where we are looking for the {\it global optimum} 
out of a large number of possibilities in a non-convex landscape (one with many saddle points, local minima/maxima, in addition to the global minimum), we cannot guarantee that the solution we find is the global optimum, namely we cannot check in 
polynomial time if what we find is indeed the optimum. (This would require comparing the solution to all other local minima, one by one, till we exhaust them all. This defies the purpose.) 

Therefore, for optimization problems in the NP-hard class, when the SOLCs reach an equilibrium, the only thing we can say is that the equilibrium found 
is the {\it best approximation} to the optimum the machine has found within an assigned time. For example, in the famous maximum satisfiability (Max-SAT) problem~\cite{computational_complexity_book}, one is given a Boolean formula 
in conjunctive normal form (Boolean variables related by OR clauses, with the different clauses related by AND gates), and asked to find the assignment of all the variables in the formula that maximizes the number of satisfied clauses (that have a truth value of 1). In general, there is no way to know that, once an assignment is found, it is indeed the maximum.  

\section{Examples}\label{example}

{\it Optimization problems $-$ } After these important considerations, we can now discuss the application of DMMs to specific hard problems. We have already shown that the simulations of SOLCs perform orders of magnitude better than the winners of the 2016 Max-SAT competition~\cite{MAXSAT_competition} on a wide variety of optimization problems. For instance, in Ref.~\onlinecite{exponential2017speedup} we have shown that the simulations of the equations of motion of SOLCs using a sequential MatLab code already offer substantial advantages over traditional algorithms for the Random 2 Max-SAT, the Max-Cut, the Forced Random Binary problem, and the Max-Clique~\cite{complexity_bible}. In some cases, the memcomputing approach finds the solution to the problem when the winners of the 2016 Max-SAT competition could not~\cite{exponential2017speedup}. 

We have also performed scalability tests on hard instances of the Max-SAT problem whose conjunctive normal form representation contains 
exactly $k$ literals ($k \geq 2$) per clause. This problem, known as Max-E$k$SAT~\cite{computational_complexity_book}, has an inapproximability gap~\cite{Hastad2001}, meaning that no known 
algorithm can overcome, in polynomial time, a fraction of the optimal solution, assuming P$\neq$NP. We have shown instead that for the hard cases considered, the simulations of SOLCs 
succeed in overcoming that gap~\cite{exponential2017speedup}. In addition, we have shown that for the SOLCs we have used in those simulations, this happens in {\it linear} time, namely the time it takes to overcome 
the inapproximability gap scales linearly with the number of variables (clauses) in the problem, vs. the exponential scalability of the best solvers of the 2016 Max-SAT competition specifically designed to tackle those problems.
\begin{figure}
	\centerline{
		\includegraphics[width=1.2\columnwidth]{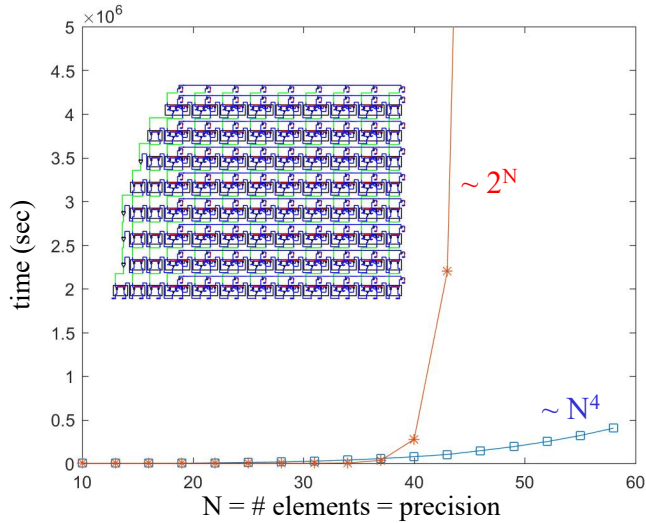}}
	\caption{\label{figSSP} Simulations of the SOLCs' differential equations, as reported in Ref.~\cite{dmm2}, describing the subset-sum problem. We consider 
		only the worst cases, namely when the number, $N$, of elements in the set is equal to the precision (number of bits), $p$, per element. For these 
		cases, there is no known pseudo-polynomial algorithm available, and the standard algorithm scales exponentially as shown by the red stars. The red curve is a guide 
		for the eye. The 
		simulations of digital memcomputing machines scale instead as a polynomial of the fourth power (blue squares and curve). The inset shows an 
		example of a small SOLC (for $N=p=9$) used to represent this problem. Both the standard algorithm and simulations of SOLCs have been implemented in a sequential MATLAB code running on a single processor of the Comet cluster of the San Diego Supercomputer Center.}
\end{figure} 

{\it Application to machine learning $-$ } In order to explore the advantages of DMMs in hard problems even further, we have also used them in the training of deep belief networks~\cite{AcceleratingDL}. In particular, we have applied them to the training of restricted Boltzmann machines (RBMs) that are difficult to pre-train. In fact, the standard way to training these networks relies  on Gibbs sampling~\cite{2018arXiv180102567R}, which is very inefficient. Quantum machines, such as D-Wave ones~\cite{Adachi2015} have been shown to substantially accelerate such 
pre-training in hardware. 

The pre-training of RBMs can be cast in the form of a quadratic unconstrained binary optimization (QUBO) problem~\cite{computational_complexity_book}. This is equivalent to finding the ground state 
(lowest energy state) of a non-convex energy landscape. By employing our memcomputing approach we have shown that the simulations of SOLCs sample very 
effectively the vast phase space defined by the energy cost function of the neural network, and provide a very good  
approximation close to the optimum~\cite{AcceleratingDL}. In fact, the acceleration of the pre-training 
achieved by {\it simulating} SOLCs is comparable to, in 
number of iterations, the reported {\it hardware} application of the quantum annealing method implemented by the D-Wave machine on the same network and data set~\cite{Adachi2015}. However, the memcomputing approach performs far better than the quantum annealing approach in terms of {\it quality} of the training~\cite{AcceleratingDL}. 

In addition, the pre-training offered by simulating SOLCs maintains a considerable advantage over other supervised learning methods, such as batch-normalization~\cite{pmlr-v37-ioffe15} and rectifiers~\cite{nair2010rectified}, that have been designed to reduce the advantage of pre-training all together. Instead, we find that the memcomputing method still maintains a quality advantage ($>1\%$ in accuracy, corresponding to a $20\%$ reduction in error rate) 
over these approaches~\cite{AcceleratingDL}. The substantial advantage of the memcomputing approach again rests on finding a very good approximation to the optimum 
compared to the traditional approaches. 

{\it A combinatorial problem $-$ } All the above cases pertain to NP-hard problems. Here, we report instead the solution of an NP-complete one, for which checking the solution is (polynomially) easy. 
Let us then consider again the subset-sum problem of Sec.~\ref{MathSOLCs}. For this problem, it is easy to choose the hardest cases: those correspond to the number of elements, $N$, in the set equal to the precision (in number of bits), $p$, required to represent each element. For $N=p$ no pseudo-polynomial algorithm is known.

The standard algorithm for these hard cases requires checking the sums of all possible subsets of the given set, till one is found that sums to the given integer $s$. 
A brute force implementation of this algorithm then diverges exponentially for these hard cases, whether the load of the calculation is done by directly checking all subsets (hence the algorithm scales as $2^N$), or half of the load is transferred to memory by storing partial sums of the computation (so that the computation scales as $2^{N/2}$ and the memory scales as $2(p+\log_{2} p)2^{N/2}$) \cite{Horowitz1974}. Either way, for these hard cases, $N=p$, the computation and/or the memory requirements diverge exponentially. 
This is evident in Fig.~\ref{figSSP}, where we show the exponential divergence of the standard algorithm with just brute force computation (no storing of 
partial results in memory). In the same figure, 
we report {\it simulations} of SOLCs representing the same instances. 

The simulations have been done by employing a {\it sequential} MATLAB implementation 
of the equations of motion of the SOLCs (see Ref.~\cite{dmm2} for the actual equations) on a single processor of the Comet cluster 
of the San Diego Supercomputer Center, so no 
parallelization has been employed. Since we solve equations of motion of the type (\ref{ODE}), the memory requirement for these simulations scales quadratically with the size of the problem, namely with $Np=N^2=p^2$. For comparison, the largest case we have considered, if done with the standard algorithm and implemented in MatLab and run on the same processor, would require more than 2,000 years to find the solution. If we used the method of storing partial-sum results in memory, we could reduce the time of computation considerably, but at an exponential cost of storage needed. 
The SOLCs' simulations instead scale as a polynomial of $\sim$4th power for that particular circuit we have chosen (a circuit example is shown in the inset of Fig.~\ref{figSSP} for a small case, $N=p=9$). The fourth 
power is easy to understand for this particular circuit realization: the circuit is spatially quadratic, and we used implicit methods to integrate forward the equations of motion. 

\section{Topology, instantons, and oracles}\label{Topology}

The above problems show that once the dynamics are initiated at some (arbitrary) point in the phase space, the system is ``guided'' towards the solution (if this is an NP-complete problem) or towards a very good approximation of the optimum, or, possibly, the optimum itself (if this is an optimization problem of a non-convex nature). It seems as though the trajectory in the (enormous) phase space is {\it constrained}, and only very specific paths are chosen to slice through this vast space to 
reach equilibria in 
an efficient way. Put differently, it is as if the physical system has some {\it global} ``knowledge'' of the whole phase space. 

Typically, global features of a space are associated to its {\it topology}, as opposed to some {\it local} (geometrical) properties of the same~\cite{Topo-book}. Therefore, 
it seems natural to ask if the DMMs we have introduced, and in particular their SOLC realization, exhibit some topological character when they attempt to find 
the equilibrium points of the dynamics. If the answer to this question is affirmative, then we have an added benefit: our machines {\it must} be robust against 
noise and structural disorder. This is because, if the search of the equilibrium points is carried out by topological objects, then in order to destroy them one 
needs to destroy the global (topological) structure of the phase space~\cite{Topo-book}. In practice, one needs to change the physical system all together.

{\it The SOLCs' architecture $-$ } A first, quite obvious, topological feature of SOLCs is the {\it architecture} of the Boolean circuits they represent. Say, we want to solve for the factorization. 
We then build the corresponding Boolean circuit and solve it ``in reverse''. We know that, given that circuit, {\it there is} a solution to the prime factorization problem. We have also designed the corresponding SOLC so that no equilibrium points may exist other than those representing the solution(s) to this problem (see 
SOLC's properties in Sec.~\ref{MathSOLCs}). 
Simply put, the solution is {\it embedded} in the architecture of the circuit. It just needs to be found. (The information embedded in the architecture of SOLCs is 
what we called {\it information overhead} in Ref.~\onlinecite{dmm2}.)

This, by itself, constrains the dynamics enormously, similar to the constrained dynamics of a liquid if it is forced to go through a maze~\cite{11_maze}: although the maze 
may be very complex, and have multiple paths, to find the exit the liquid is constrained to go through only those paths that solve the maze. The architecture (topology) of the maze already forces the liquid to follow specific paths, even if the corresponding phase space of the dynamical system may be large. 

In a similar vein, the architecture of the SOLCs forces the ``electron liquid'' (if we now interpret those circuits as actual electronic components) to go through 
specific paths in the phase space, till the liquid reaches the equilibrium points (the exit points of the ``maze''). However, the architecture of the circuit is 
not the only topological feature of the physical systems representing DMMs. 

{\it The topology of phase space $-$ } A vast phase space with a non-convex landscape supports {\it critical points}, namely points, $\xvec_{cr}$, at which the flow vector 
field $F$ in Eq.~(\ref{ODE}) is zero~\cite{perko_01}: 
	\begin{equation}
F(\xvec_{cr})=0\label{fixed}.
\end{equation}
In a neighborhood of any critical point we can perform linear stability analysis, namely we can expand Eq.~(\ref{ODE}) to linear order:
\begin{equation}
\dot{ {\bf x}} \approx {\bf J}({\bf x}_{cr})({\bf x} - {\bf x}_{cr}),\label{local-cr}
\end{equation}
where $J$ is the Jacobian matrix
\begin{equation}
[J({\bf x})]_{ij} = \frac{\partial F_i({\bf x})}{\partial x_j}.
\end{equation} 
We can now determine the eigenvalues, $\lambda_i$, and eigenvectors, ${\bf v}_i$, of this matrix for the given critical point. 
The linearized equation~(\ref{local-cr}) then results in the local trajectories
\begin{equation}
{\bf x}(t) \approx {\bf x}_{cr} + \sum_i {\bf v}_i e^{\lambda_i t}\label{approx-cr}.
\end{equation}

The eigenvectors corresponding to $\text{Re } \lambda_i < 0$ and $\text{Re }\lambda_i > 0$ define the vector spaces tangent to the {\it stable} and {\it unstable manifolds}, respectively, at each critical point. For a negative eigenvalue, from Eq.~(\ref{approx-cr}), it is clear that the dynamics tend to bring the system back to the critical point; that direction is {\it attractive} (stable direction). Instead, a positive eigenvalue indicates that the 
corresponding direction in the phase space is {\it repulsive} (unstable direction). Those critical points that have some attractive and some repulsive directions are saddle points, while those having only attractive (repulsive) directions are local minima (maxima)~\cite{perko_01}. In fact, there exist both strong (statistical mechanics) arguments~\cite{fyodorov, dean} and empirical evidence~\cite{non-convex} suggesting that in non-convex landscapes, the number of local minima is exponentially smaller than the number of saddle points with increasing size of the phase space, which renders standard local methods for optimization, such as gradient descent, inefficient~\cite{non-convex}.

All eigenvectors with $\text{Re }\lambda_i= 0$ span a flat manifold tangent to a \emph{center manifold}~\cite{perko_01}. It is clear from Eq.~(\ref{approx-cr}) that center manifolds are somewhat irrelevant to the dynamics, since in those manifolds the system does not ``move''. 

The critical points are of topological character, namely their number and {\it index} (the number of their unstable directions) is only determined by the topology of the phase space~\cite{Fomenko}. Although it is not so easy to determine the number and index of critical points, especially when the dimension of the phase space is large, in Ref.~\onlinecite{Bearden} we have done an extensive search of critical points for a simplified version of AND and OR SOLGs. That analysis has shed a lot of light about the 
dynamics of these systems. 

It turns out that the unstable critical points are such that the {\it magnitude} of the eigenvalues, $\lambda_i$, of their unstable directions, is much smaller than the magnitude of the eigenvalues of their stable directions~\cite{Bearden}. This is consistent with other numerical results showing that generic high-dimensional non-convex landscapes have mainly saddle points surrounded by quasi-flat unstable directions~\cite{non-convex}. The amount of memory in the system then determines the degree of curvature of the unstable directions. This means that the system, starting from an arbitrary initial 
condition in the phase space, can easily get ``very close'' to these critical points, despite having repulsive (unstable) directions. 
Once the system has reached a critical point with some unstable directions, it can get out of it and move somewhere else. 

{\it Instantons $-$ } Here, another topological feature emerges. Instead of wandering around the vast phase space aimlessly, the system can take advantage of particular trajectories that connect critical points with {\it different} indexes. These trajectories are called {\it instantons}~\cite{Coleman}, and they are the classical (Euclidean) analogue of quantum tunneling, and, apart from the ``instant'' at which ``tunneling'' occurs, the system spends most of its time at the critical points ({\it classical vacua})~\cite{Schwarz}, with this time determined again by the amount of memory in the system. Memory that is too large or too small may lead to a considerable slowing down of the dynamics~\cite{Bearden}. 

Instantons are topologically non-trivial deterministic trajectories, $\xvec_{cl}$, (namely functions that have different limits for $t\rightarrow -\infty$ and $t\rightarrow +\infty$) satisfying~\cite{Schwarz}:
\begin{eqnarray}
\dot \xvec_{cl}(t,\sigma) = F(\xvec_{cl}(t,\sigma));\;\;\;\;\;\;  \xvec_{cl}(\pm\infty,\sigma)=\xvec_{\textrm cr}^{i,f},\label{instanton}
\end{eqnarray}
with $\xvec_{\textrm cr}^i$ and $\xvec_{\textrm cr}^f$ arbitrary critical points of the flow vector field $F$ (of Eq.~(\ref{ODE})) with different indexes. The parameters $\sigma$ are the so-called modulii of instantons, which can be used as their coordinates, and represent the {\it non-local} character of instantons~\cite{Coleman}. 

Why would a dynamical system employ these instantons in the first place? This is because, if one writes Eq.~(\ref{ODE}) in a path-integral representation~\cite{topo}, 
the partition function of such representation has an action functional, $S_{\textrm {Eucl}}$, which is of topological character~\cite{Entropy}. As any physical system, the trajectory chosen 
by the system is the one that renders such action functional {\it stationary}~\cite{Goldstein}. Instantons turn out to be those trajectories that render the topological $S_{\textrm {Eucl}}$ stationary~\cite{Coleman,Schwarz}
\begin{equation}
\delta S_{\textrm {Eucl}}(x_{cl}(t,\sigma))=0.
\end{equation}

In reality, instantons define a {\it family} of classical trajectories (all related to each other via some symmetry 
transformation~\cite{Book1}), namely there may be more than one path rendering the action functional stationary. In addition, they are present because the equations representing SOLCs are non-linear, and they emerge 
only during the {\it transient} dynamics, namely before the system has reached a steady state~\cite{topo}. 

\begin{figure}[t!]
	\centerline{
		\includegraphics[width=1.2\columnwidth]{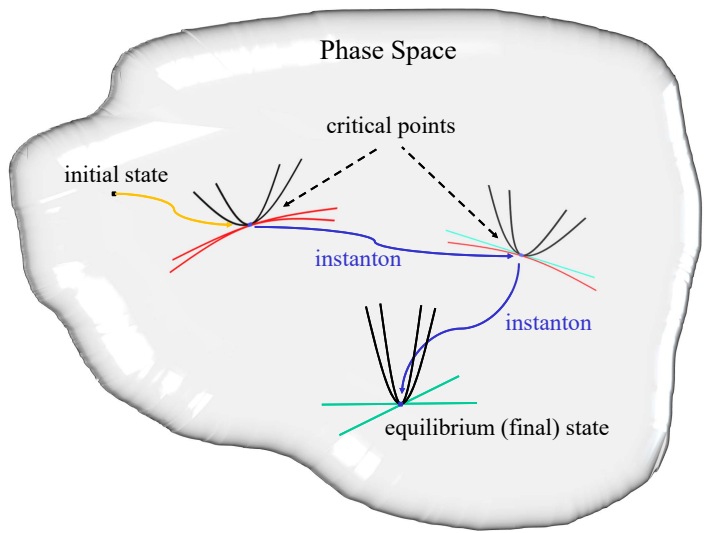}}
	\caption{\label{Instantons} Schematic of the microscopic dynamics in the phase space of a SOLC representing a particular problem. Starting from an 
	arbitrary initial condition, the system ``falls'' into the closest critical point with some unstable directions (red parabolas) and stable directions (black parabolas), a saddle point. The curvature of the unstable directions is much smaller, in magnitude, than the curvature of the stable directions. The system then evolves to another critical point (saddle point) with lower index (less unstable directions) along an instanton trajectory. This step is repeated till the system ends up into a fully stable equilibrium point. Along the way, the unstable directions of the critical points have evolved into center manifolds (green straight lines) representing the memory degrees of freedom.}
\end{figure}

The microscopic dynamics of a SOLC then proceeds as follows (see Fig.~\ref{Instantons})~\cite{Bearden}: the system starts from an arbitrary initial condition in the phase space, which is not necessarily a critical point. It is then attracted by the closest unstable critical point (which is not unstable enough to repel the dynamics). After the system reaches the first critical point, it can only go through instantonic 
trajectories to make the action $S_{\textrm {Eucl}}$ stationary. 

Therefore, the system ``hops'' from one critical point (saddle point) to another with lower index (namely more 
stable), until it reaches the last critical point, the equilibrium of the dynamics, which has only stable directions, and possibly center manifolds. In fact, in doing so, the unstable directions have become center manifolds. The latter ones represent  
the arbitrariness of the internal state variables when the system has reached equilibrium~\cite{dmm2}. This is because, the SOLGs (and corresponding SOLCs) have 
been designed in such a way that they satisfy their logical proposition in either voltages or currents~\cite{dmm2}. The internal state variables, $\tilde x$, in Eqs.~(\ref{Geq1}) and~(\ref{Geq2}), representing memory, need only to provide the extra degrees of freedom for the system to self-organize into the correct solution. Once the latter has been reached, the internal state variables lose their purpose, and the equilibrium is stable irrespective of the values of those variables. This is why, 
at equilibrium, the directions associated with the internal state variables define center manifolds in the phase space. 

Note that the number of unstable directions can be at most equal to the number of state variables. These, in turn, grow only {\it polynomially} with the size of the corresponding Boolean problem. Therefore, even if each instanton connects critical points differing by only one unstable direction, the number of ``instantonic steps'' necessary to reach equilibrium can only be equal to or less than the number of state variables, namely the dynamics employ a number of instantonic steps growing at most {\it polynomially} with the size of the problem~\cite{dmm2}. This is another (topological) way of understanding the polynomial requirements of SOLCs in solving hard problems. 

The transient (instantonic) phase of the dynamics is reminiscent of an {\it avalanche} phenomenon, in which ``energy'' is released in steps till the lowest ``energy'' is reached \cite{Lu1995}. This analogy is not far fetched since it has been shown that any classical dynamical system (with and without noise) can be expressed within a topological field theory~\cite{Entropy}. The topological field theory that emerges is of a Witten type~\cite{Witten1}. Therefore, any transient dynamics can be described within the same instantonic 
formalism, where, of course, the critical points, and the topological sector of the theory change according to the physical system considered. 

{\it Long-range order $-$ } There is yet another important feature of the instantonic (transient) phase worth stressing. If one computes matrix elements of certain (topological) observables 
on instantons, one finds that those matrix elements are {\it independent} of both space {\it and} time (they are {\it topological invariants})~\cite{topo}. This means that the ``tunneling'' between critical points 
along instantonic trajectories displays an {\it ideal long-range order}, similar to entanglement in Quantum Mechanics. However, note a fundamental difference between the instantonic long-range order and entanglement. In the latter case, the quantum entangled state is prepared by the experimentalist at the outset, namely it is there at the beginning of the (entangled) quantum dynamics. In the instantonic case, it takes some time for the system to reach the first critical point from an arbitrary initial condition, namely it takes some time for the system to reach a ``rigid state''. In addition, quantum entanglement is destroyed by decoherence effects, an unwanted eventuality. Instead, the long-range order of the instantonic dynamics, once established, cannot be destroyed by noise or perturbations (see also below)~\cite{topo}. It naturally disappears once the system has reached equilibrium. 

For SOLCs, one type of 
topological observable is precisely the one that ``detects'' when voltages at every terminal in the SOLC switch from a logical 1 to a logical 0, and vice versa~\cite{topo}. This is the observable that is directly related to the {\it correlations} between voltages at different terminals anywhere in the circuit.  Therefore, if the instanton matrix elements on these topological observables are independent of space and time, the correlations between voltages at different terminals anywhere in the circuit must share a similar feature, which is what is found both analytically and numerically~\cite{topo}.

This is precisely the property we were 
after: the correlations between voltages of all SOLGs 
of the Boolean circuit representing a problem {\it do not} decay spatially. Therefore, if a gate needs to satisfy its logical proposition, it can do so by ``exchanging'' truth values with 
another gate {\it arbitrarily far away}, despite the connections in the Boolean circuit representing the problem are {\it local}. The system becomes {\it rigid}. This {\it collective} dynamical behavior is reminiscent of a phase transition~\cite{Goldenfeld}, or even the ``edge of chaos''~\cite{edge-chaos}. However, unlike those cases, where 
the spatial correlations are scale-free (they decay as a power law), in SOLCs the correlations do not decay at all~\cite{topo}.

This last point explains why SOLCs can efficiently tackle hard, non-convex problems, while standard algorithms (relying only on {\it local} information) cannot. In the worst cases, non-convex problems expressed in Boolean form as, e.g., conjunctions of disjunctive clauses between variables, are such that when a certain amount of satisfied clauses is reached, any further improvement requires many {\it simultaneous/correlated} flips of variables (changing their values from 0 to 1 or 
vice versa). This is a {\it non-local} (global) type of assignment~\cite{computational_complexity_book}. 
Therefore, without a {\it global} knowledge of the solution space, a standard combinatorial algorithm is bounded to explore a vast number of possibilities, requiring exponential 
resources to do so (which leads to the inapproximability gap~\cite{Hastad2001} previously mentioned). Instead, by employing instantons, hence fundamentally non-local objects, a dynamical system representing the same Boolean problem can easily correlate variables {\it anywhere} they appear in the problem specification.   

The independence on time of the instantonic matrix elements on topological observables represents instead a strong dependence of the {\it trajectories} connecting 
critical points on initial conditions (but {\it not} of the critical points themselves). This means that the system can follow completely different trajectories to find the equilibrium 
points in the phase space, according to the initial conditions assigned. This, however, does not change the robustness of the solution search due to the topological 
character of the critical points. The latter indeed cannot be destroyed or changed even if one explicitly adds noise to the internal state variables, unless that 
noise literally destroys the physical system itself. The ensuing robustness of SOLCs with respect to noise and structural disorder was 
demonstrated explicitly in Ref.~\cite{Bearden}.

{\it Instantons as oracles $-$ } Finally, let us make another interesting consideration. We have discussed that once a Boolean problem is expressed as a dynamical system in terms of SOLGs, the 
system is guided towards the solution, and does so following specific trajectories (instantons) connecting critical points of increasing 
stability in the phase space. Due to the gigantic 
size of the phase space (even for relatively ``small'' Boolean problems), there is {\it no way} for us to know {\it a priori} which path the system will take, 
once an initial condition is assigned. The 
physical system is too {\it complex} for us to infer its collective properties from its elementary constituents (this is the very definition of a complex system). 

Therefore, even if the systems we 
consider are {\it deterministic}, their complexity is such that their actual {\it internal workings}, which determine the dynamical paths that connect input and  output, are {\it not known} down to their microscopic details at every instant of time. We know which path the machine has taken only {\it after} we have observed it (or calculated it). A machine of this type is what is known in computer science as an {\it oracle}~\cite{computational_complexity_book}. In fact, an oracle is one that, via some internal mechanism unknown to the user, is able to correctly ``guess'' the next step of a computation done by a Turing machine, and guide that machine to the correct solution 
in polynomial time if the solution tree grows exponentially with problem size~\cite{computational_complexity_book}. In a similar vein, instantons are able to ``guess'' correctly the path they have to travel in the phase space. Their ``guess'' is of course guided 
by physical laws of motion and the principle of stationary action. However, to an external observer, it would be difficult, if not outright impossible to know 
in advance the path they will take. We can then interpret the instantonic phase of DMMs as a physical realization of an oracle. \\

\section {Conclusions}

In this perspective we have provided an overview of {\it digital} (hence scalable) memcomputing machines~\cite{dmm2}, their practical realization using a new type of logic framework (self-organizing logic), and the Physics behind their dynamics. We have discussed that their dynamical behavior proceeds via instantons: families of classical trajectories in the phase space that connect critical points of increasing stability. The topological character of these objects renders these machines robust against noise and structural 
disorder. In addition, the non-locality of instantons generates an ideal long-range order reminiscent of entanglement in Quantum Mechanics. It is this long-range 
order that allows these physical machines to correlate gates at arbitrary distance, hence allowing an efficient non-local (global) search in the solution space of the problem they are 
designed to solve. 

Since the equations of motion describing the dynamics of these non-quantum systems are just coupled ordinary differential equations, they can be efficiently simulated in software using our modern computers. In addition, if implemented using electronic components, these machines can be built within CMOS technology offering a 
path to real-time computing for several applications of current interest, such as machine learning \cite{Witten2016}, autonomous vehicles \cite{litman2017autonomous}, robotics \cite{Nolfi2000}, etc. 

Software simulations of these machines have already shown a considerable advantage over traditional (algorithmic) approaches.  
We have referred to several examples regarding non-convex optimization problems that have already appeared in the literature, where these advantages are 
particularly evident. In this perspective we have also shown the solution of hard instances of a combinatorial problem, the subset-sum, showing that simulations 
of these memcomputing machines offer solutions to this problem in polynomial time vs. the exponential requirements of standard approaches.  

The problems we have tackled so far have been restricted to the (albeit vast) space of Boolean problems for which these machines are ideally suited. However, there are several other problems that require continuous variables. The most obvious way, although possibly not the most efficient, to tackle these problems is to discretize the continuous variables into an approximate binary representation, and then apply the same machines we have discussed here to that representation~\cite{manukian2017inversion}. This is how these problems are represented even in our modern computers. However, we believe there is room for improvement if the self-organizing 
gates are directly modified to account for such continuous-variable instances. 

Another interesting area of research is to apply these machines to efficiently finding the ground state 
(or spectrum) of {\it quantum} Hamiltonians. Finding the ground state (spectrum) of quantum Hamiltonians would have tremendous consequences both for fundamental science, as well in disparate practical 
applications, such as drug discovery and biotechnology \cite{Raha2007,Chodera2011,Wang2015}, design of materials with desired properties \cite{Council2003}, etc. Universal memcomputing 
machines have been shown to be quantum-complete, meaning that they can, in principle, simulate quantum machines~\cite{ONUMM}. However, that result is merely theoretical, 
and does not specify the amount of resources required for such a simulation. Therefore, an efficient mapping between the quantum problem and a non-quantum Boolean solution needs to be found. Note that, in view of what we have discussed, it is enough for us to find a mapping that transforms the hard optimization problem of finding the ground state of a quatum Hamiltonian into an equally hard Boolean problem. More work in this direction is thus necessary.

Irrespective, we have already shown that there exist engineered dynamical systems that combine the power of {\it physical} processes with the {\it digital} structure 
of input and output, thus allowing the design of machines that are easily {\it scalable}. Once more, as in the case of quantum computing where one may 
efficiently factor numbers on a quantum computer, physics-based approaches to computation seem to offer benefits that are difficult, if not impossible to obtain from 
only traditional, algorithmic approaches. We then hope this work will motivate further research in this promising area of computing with Physics, in particular 
memcomputing. 

\emph{Acknowledgments $-$} We thank Yuriy Pershin, Forrest Sheldon, Haik Manukian, Sean Bearden, and Yan Ru Pei for useful discussions and for their contribution over the years on various aspects of memcomputing. We also thank Dr. Pietro Cicotti of the San Diego Supercomputer Center (SDSC) for allowing us to publish the data reported in Fig.~\ref{figSSP} on the subset-sum problem that he has generated using a 
sequential MATLAB code running on a single processor of the Comet cluster of the San Diego Supercomputer Center. M.D. acknowledges partial support from the Center for Memory and Recording Research at UCSD.

%


\end{document}